\documentclass[12pt]{article}
\begin{document}
\newcommand {\sheptitle}
{ Thermal and non-thermal leptogenesis in different neutrino mass models  with tribimaximal mixings}
\newcommand{\shepauthor}
{  N. Nimai  Singh $^{\dag,*}$\footnote{Regular Associate of
    ICTP.\\{\it{E-mail address}}:nimai03@yahoo.com}, H. Zeen  Devi$^{\dag}$     and  Amal Kr
  Sarma$^{\ddag}$  }
\newcommand{\shepaddress}
{$^{\dag}$ Department of Physics,Gauhati University,Guwahati-781014,India \\
$^{\ddag}$ Department of Physics, D.R College, Golaghat,India\\
$^*$The Abdus Salam International Centre for Theoretical Physics,
Strada Costiera 11, 31014 Trieste, Italy. }
\newcommand{\shepabstract}
 {In the present work we study both thermal and non-thermal
   leptogenesis in all neutrino mass models describing the presently available
   neutrino mass patterns. We consider the Majorana CP violating
   phases coming  from
   right-handed Majorana mass matrices to estimate the baryon
   asymmetry of the universe, for
   different neutrino mass models namely 
 degenerate,  inverted hierarchical and normal hierarchical models, with
 tribimaximal mixings. Considering two possible  diagonal forms of 
  Dirac neutrino mass matrix as either charged lepton or up-quark mass
  matrix, the right-handed Majorana
 mass matrices are constructed from the
 light neutrino mass matrix through the inverse seesaw formula. Only the
 normal hierarchical model leads to the best  predictions
 for baryon asymmetry of the universe, consistent with observations in both thermal and
 non-thermal leptogenesis scenario. The analysis though phenomenological may
 serve as an additional information in the discrimination among the
 presently available 
 neutrino mass models.} 

\begin{titlepage}
\begin{flushright}
\end{flushright}
\begin{center}
{\large{\bf\sheptitle}}
\bigskip \\
\shepauthor
\\
\mbox{}
\\
{\it \shepaddress}
\\
\vspace{.5in}
{\bf Abstract} \bigskip \end{center}\setcounter{page}{0}
\shepabstract
\end{titlepage}
\section{Introduction}
The existence of heavy right-handed Majorana  neutrinos in some of the
left-right symmetric GUT models,  not only gives small but non vanishing neutrino masses through the
 celebrated seesaw mechanism[1],  it also  plays an important role
in explaining the baryon asymmetry
 of the universe [2] $ Y_{B} $ =$(6.1_{-0.2}^{+0.3})$ $10^{-10}$. 
Such an asymmetry can be dynamically generated if the particle
interaction rate and the  expansion rate
 of the universe satisfy Sakharov's three famous conditions [3].
 Majorana right-handed neutrinos satisfy the second
 condition i.e, C and CP violation as they can have an
 asymmetric decay to leptons and Higgs particles, and the
 process occurs at different rates for particles and
 antiparticles. The lepton asymmetry is then partially converted to
 baryon asymmetry through the non-perturbative electroweak sphaleron
 effects [4,5]. In such thermal leptogenesis the right-handed
 neutrinos can be generated thermally after inflation, if their masses
 are comparable to or below the reheating temperature $M_1\leq T_R$. This
 allows high scale reheating temperature $T_R\geq 10^9$ GeV[ 6]. In
 non-thermal leptogenesis[7] it is possible to produce lepton asymmetry by
 using the low reheating temperature, where the right-handed neutrinos
 are produced through the direct non-thermal decays of the inflaton.  
 This is particularly important for
 supersymmetric models where gravitino problem[8] can be avoided provided
  the reheating temperature after inflation is bounded from above
 in a certain way, namely $T_R\leq (10^6 - 10^7)$ GeV.

  In order to calculate the baryon asymmetry from a given neutrino
mass model,  one usually  starts with the light 
neutrino mass matrices $m_{LL}$ and then  relates it with  the heavy
 Majorana neutrinos $M_{RR}$  and
 the Dirac neutrino mass  matrix $m_{LR}$  
through inverse seesaw mechanism in an elegant way. We consider  the Dirac neutrino
mass  matrix $m_{LR}$ as either the  charged 
lepton mass matrix or up quark mass matrix for phenomenological analysis. The complex CP violating
phases are usually derived from the MNS leptonic
mixing matrix. In the present work we are interested to  consider the
complex Majorana phases which are derived from
 the right-handed Majorana mass matrix $M_{RR}$, in the estimation of 
  baryon asymmetry of the universe. We wish to consider
 the left-handed light Majorana neutrino mass matrices $ m_{LL}$ which obey the
 $\mu-\tau$ symmetry[9]  where tribimaximal mixings[10] are realised, for
all possible patterns of neutrino masses, viz, degenerate, inverted hierarchical and normal hierarchical mass
 patterns.  We first parametrise the light left-handed Majorana neutrino mass
 matrices which are subjected to correct predictions of neutrino mass
 parameters and mixing angles.  The calculation of baryon asymmetry of
 the universe in the light of thermal as well as non-thermal
 leptogenesis, may serve as an
 additional information to further discriminate the correct pattern of
 neutrino mass models and also shed light on the structure of Dirac
 neutrino mass matrix.  
 
In section 2, we briefly mention the formalism for estimating
the lepton asymmetry in thermal leptogenesis  through out of the equilibrium 
decay of the heavy right-handed Majorana neutrinos, followed by  numerical
calculation and results. Section 3 is devoted to non-thermal
leptogenesis and numerical predictions. Finally in section 4 we
conclude with  a  summary and discussion. Important expressions related to $m_{LL}$
which obey $\mu-\tau$ symmetry for three neutrino mass models, are
relegated to Appendix A.

\section{ Baryon asymmetry of the universe in thermal leptogenesis}
The canonical seesaw formula (known as type-I)[1] relates the left-handed Majorana  neutrino mass matrix $m_{LL}$ 
and heavy right handed Majorana mass matrix $M_{RR}$ in a  simple way
\begin{equation}\label{ch301}
m_{LL} = -m_{LR}M_{RR}^{-1}m_{LR}^{T}
\end{equation}
where  $m_{LR}$ is the Dirac neutrino mass matrix. 
For our calculation of lepton asymmetry, we consider the model[5,11]
 where the asymmetric decay of the lightest of the heavy right-handed Majorana neutrinos,
 is assumed.  The physical  Majorana neutrino $N_{R}$  decays into two modes:
\begin{center}
$N_{R}\rightarrow l_{L}+\phi^{\dagger}$\\

      $\rightarrow \overline{l}_{L}+\phi$
\end{center}
where  $l_{L}$ is the lepton and $\bar{l}_{L}$ is the antilepton and  the
 branching ratio for these two decay modes is  likely to be different. 
The CP-asymmetry which is caused by the intereference of tree level with one-loop corrections
 for the decays of lightest of heavy right-handed Majorana neutrino $N_{1}$, is defined by [5,12]
\begin{center}
$\epsilon=\frac{\Gamma -\overline{\Gamma}}{\Gamma+\overline{\Gamma}}$
\end{center}
where  $\Gamma=\Gamma(N_{1}\rightarrow l_{L}\phi^{\dagger})$ 
and $\overline{\Gamma}=\Gamma(N_{1}\rightarrow \overline{l_{L}}\phi)$ are the decay rates.
A perturbative calculation from the interference between
 tree level and vertex plus self energy diagrams, gives[13] the lepton asymmetry
 $\epsilon_1$ for  non-SUSY case as
\begin{equation}
\epsilon_i=-\frac{1}{8\pi} \frac{1}{(h^{\dagger}h)_{ii}} \sum_{j=2,3}
Im[(h^{\dagger}h)_{ij}]^2 [f(\frac{M^2_j}{M^2_i})+
g(\frac{M^2_j}{M^2_i})]
\end{equation}
where $f(x)$ and $g(x)$ represent the contributions from vertex and
self-energy corrections respectively,
$$f(x)=\sqrt{x}[-1+(x+1)ln(1+\frac{1}{x})],$$
$$g(x)=\frac{\sqrt{x}}{x-1}.$$
For hierarchical right-handed neutrino masses where $x$ is large, we
have the approximation[2], $f(x)+g(x)\simeq \frac{3}{2\sqrt{x}}.$
This simplifies to 
\begin{equation}\label{ch305}
 \epsilon_{1} \simeq
-\frac{3}{16\pi}\left[\frac{Im[(h^{\dag}h)^{2}_{12}]}{(h^{\dag}h)_{11}}\frac{M_{1}}{M_{2}}
  +
 \frac{Im[(h^{\dag}h)^{2}_{13}]}{(h^{\dag}h)_{11}}\frac{M_{1}}{M_{3}}\right]
\end{equation}  
where  $h=m_{LR}/v$ is the Yukawa coupling of the Dirac neutrino mass
matrix in the diagonal basis of $M_{RR}$.
In term of light Majorana neutrino mass matrix $m_{LL}$, the above
expression can be simplified to 
$$ \epsilon_1\simeq - \frac{3}{16\pi} \frac{M_1}{(h^{\dagger}h)_{11}v^2}
Im(h^{\dagger}m_{LL}h^*)_{11}.$$
For quasi-degenerate spectrum i.e., for  $M_1\simeq M_2< M_3$ the
asymmetry is largely  enhanced by a resonance factor and in such situation, the lepton asymmetry is modified[14] to
\begin{equation}\label{ch306}
 \epsilon_{1} \simeq \frac{1}{8\pi}\frac{Im[(h^{\dag}h)^{2}_{12}]}{(h^{\dag}h)_{11}} R 
\end{equation}
 where 
\begin{center}
$R=\frac{M_2^{2}(M_2^{2}-M_1^{2})}{(M_1^{2}-M_2^{2})^2 + \Gamma^2_{2} M_{1}^{2}}$ and $\Gamma_{2}=\frac{(h^{\dag}h)_{22} M_2}{8 \pi}$
\end{center}
It can be noted that in case of SUSY, the functions $f(x)$ and $g(x)$ are given by
$f(x)=\sqrt{x}ln(1+\frac{1}{x})$ and $g(x)=\frac{2\sqrt{x}}{x-1}$; and
for large $x$ one can have $f(x)+g(x)\simeq \frac{3}{\sqrt{x}}$. Therefore
the factor $\frac{3}{8}$ will appear  in place of $\frac{3}{16}$ in the
expression of CP asymmetry[2].

The CP asymmetry parameter $\epsilon_1$ is related to the leptonic
 asymmetry parameter through $Y_L$ as 
\begin{equation}
Y_L\equiv\frac{n_L-\bar{n_L}}{s}=\sum_{i}^{3}\frac{\epsilon_i\kappa_i}{g_{*i}}
\end{equation}
where $n_L$ is the lepton number density, $\bar{n_{L}}$ in the
anti-lepton number density, $s$ is the entropy density, $\kappa_i$ is
the dilution factor for the CP asymmetry $\epsilon_i$, and $g_{*i}$ is
the effective number of degrees of freedom at temperature $T=M_i$. The
baryon asymmetry $n_B$ produced through the sphaleron transmutation of
$Y_L$, while the quantum number $B-L$ remains conserved, is given by 
[15]
\begin{equation}
\frac{n_B}{s}=CY_{B-L}=\frac{C}{C-1}Y_L
\end{equation}
where
\begin{equation}\label{ch309}
C= \frac{8N_F+4N_{H}}{22N_F+13N_{H}}.
\end{equation}
Here $N_F$ is the number of fermion families and  $N_{H}$ is the
number of Higgs doublets. Since $s=7.04 n_{\gamma}$ the baryon number
density over photon number density $n_{\gamma}$ corresponds to the  observed baryon asymmetry  of the Universe[16],
\begin{equation}\label{ch311}
Y_B^{SM}\equiv (\frac{n_{B}}{n_{\gamma}})^{SM} \simeq  d     
  \kappa_{1} \epsilon_{1} 
\end{equation}
where $d\simeq 0.98\times
10^{-2}$ is used in the present calculation.
 In case of MSSM, there is no major numerical change with
 respect to the non-supersymmetric case in the estimation of baryon asymmetry. One expects approximate
 enhancement factor of about  $\sqrt{2} (2\sqrt{2})$ for strong (weak) washout regime[2].

In the expression for baryon-to-photon ratio $\kappa_{1}$ descrides
the washout of the lepton asymmetry 
 due to various lepton number violating processes. This efficiency
 factor (also known as dilution factor)  mainly depends on the effective neutrino mass $\tilde{m_{1}}$
\begin{center}
$\tilde{m_{1}}=\frac{(h^{\dag}h)_{11}v^2}{M_{1}}$
\end{center}
where $v$ is the electroweak vev, $v=174 GeV$.
For $10^{-2}eV<\tilde{m_{1}}<10^3eV$,  the washout factor $\kappa_{1}$ can be well approximated by[12,17]
\begin{equation}\label{ch314} 
\kappa_{1}(\tilde{m_{1}})=0.3 \left[\frac{10^{-3} }{\tilde{m_{1}}}\right]\left[log\frac{\tilde{m_{1}}}{10^{-3}}\right]^{-0.6}.
\end{equation}
We adopt a single expression for $\kappa_{1}$  valid only for the given range of $\tilde{m_{1}}$[17,18,19].
\subsection{Numerical calculations and results}
To compute the numerical results,   we first choose the light left-handed
Majorana neurino mass matrix $m_{LL}$ proposed in
Appendix A. These mass matrices obey the $\mu-\tau$ symmetry[9] which
guarantees the tribimaximal mixings[10]. The input parameters are fixed at
the stage of predictions of neutrino mass parameters and mixings given in
Table 1. These results are consistent with the recent data on neutrino
oscillations. 

For the calculation of baryon asymmetry, we then translate these mass  matrices to $M_{RR}$ via inversion of
 the  seesaw formula, $ M_{RR}=-m_{LR}^{T}m_{LL}^{-1}m_{LR}$. We
 choose a basis $U_{R}$ where $M_{RR}^{diag} = U_{R}^{T} M_{RR} U_{R}$=diag($M_{1},M_{2},M_{3}$)
 with real and positive eigenvalues.
 We then transform diagonal form of Dirac mass matix, $m_{LR}$=diag($\lambda^{m},\lambda^{n},1)v$ to the
 $U_{R}$ basis:   $m_{LR} \rightarrow m'_{LR} = m_{LR} U_{R} Q$ where
 $Q = diag ( 1,e^{i \alpha}, e^{i \beta})$ is the complex matrix containing
 CP-violating Majorana phases derived from $M_{RR}$.
 Here $\lambda$ is the Wolfeinstein
 paramater and the choice $(m,n)$ in $m_{LR}$  gives the type of Dirac mass matrix.
 For example, $(6,2)$ for
 charged-lepton type mass matrix and $(8,4)$ for up-quark type mass
 matrix. In this prime basis the
 Dirac neutrino Yukawa coupling becomes  $h^{\prime} =
 \frac{m'_{LR}}{v}$ which enters in the expression of CP-asymmetry $\epsilon_1$.
The Yukawa coupling matrix  $h^{\prime}$ 
  also becomes complex, and hence the term $Im( h^{\dag}h)_{1j}$
  appearing in lepton asymmetry $\epsilon_1$ gives a non-zero
  contribution. A straightforward simplification shows that
  $(h^{\dagger}h)^2_{1j}= (Q^*_{11})^2Q_{22}^2 R_2 +
  (Q^*_{11})^2Q_{33}^2 R_3$
 where $R_{2,3}$ are real parameters. After inserting the values of phases the above
  expression leads to 
 $Im(h^{\dagger}h)^2_{1j}=-[R_2 \sin  2(\alpha-\beta)+ R_3 \sin 2\alpha]$
 which imparts non-zero CP
  asymmetry for particular choice of $(\alpha, \beta)$.  

In our numerical estimation of lepton asymmetry, we choose some
arbitrary values of $\alpha$ and $\beta$ other than $\pi/2$ and $0$. For example,
 light neutrino masses $(m_1, -m_2, m_3)$ lead to $M_{RR}^{diag}= diag
 ( M_1, -M_2, M_3)$, and  we  thus fix the Majorana phase 
 $Q = diag ( 1,e^{(i \alpha)}, e^{(i \beta)}) = diag(1,e^{i( \pi/2 +
   \pi/4)},e^{i\pi/4})$ for $\alpha =(\pi/4 +\pi/2)$ and $\beta =
 \pi/4$. The extra phase $\pi/2$ in $\alpha$ absorbs the negative sign
 before heavy Majorana mass $M_2$. In our search programme such choice
 of the phases leads to highest numerical estimations of lepton CP
 asymmetry.

 In Table 1 we give the predictions on $\bigtriangleup m^2_{21}$ and
$\bigtriangleup m^2_{23}$ of these seven neutrino mass models under
consideration in Appendix A. They obey $\mu-\tau$ symmetry and predict
tribimaximal mixings in addition. In Table 2 the three heavy right-handed
neutrino masses are extracted from the right-handed Majorana mass
matrices so constructed through inverse seesaw formula, for three
choices of diagonal Dirac neutrino mass matrices. We get degenerate
spectrum of heavy Majorana masses for normal hierarchical model and this allows us to use
resonant leptogenesis formula. The corresponding baryon asymmetry
$Y_B$ are estimated in Table 3 and this shows that only normal
hierarchical model predict reasonable values whereas inverted
hierarchical model (IIB) nearly misses the observational
bound. Degenerate models predict too low baryon asymmetry. 


\begin{table}[tbp]
\begin{tabular}{cccccc}\hline
Type&$\Delta m^{2}_{21}[10^{-5}eV^{2}]$&$\Delta m^{2}_{23}[10^{-3}eV^{2}]$&$\tan^{2}\theta_{12}$&$\sin^{2}2\theta_{23}$&$\sin\theta_{13}$\\
\hline
 Deg.(IA) &7.8&2.6&0.5&1.0&0.0\\
 Deg.(IB) &7.9&2.5&0.5&1.0&0.0\\
 Deg.(IC) &7.9&2.5&0.5&1.0&0.0\\
\hline
 Inh.(IIA) &7.3&2.5&0.5&1.0&0.0\\
 Inh.(IIB) &8.5&2.3&0.5&1.0&0.0\\
\hline
Nh.(IIIA) &7.1&2.1&0.5&1.0&0.0\\
Nh.(IIIB) &7.5&2.4&0.5&1.0&0.0\\
\hline
\end{tabular}
\hfil
\caption{\footnotesize  Predicted values of the  solar and atmospheric neutrino
mass-squared differences  for $\tan^{2}\theta_{12}$=0.50, using  $m_{LL}$  given in the
Appendix A.   }
\end{table}


\begin{table}[tbp]
 \begin{tabular}{l l l l l} \hline
Type & (m,n) & $ M_{1}$ & $ M_{2}$ & $ M_{3}$ \\ \hline
IA & (6,2) & 1.22 $ \times$ $10^{8}$ & -6.01 $\times$ $10^{11}$ & 2.59 $\times$ $10^{13}$\\
IA & (8,4) & 9.86 $ \times$ $10^{5}$ & -5.03 $\times$ $10^{9}$ & 2.51 $\times$ $10^{13}$\\ 
IB & (6,2) & 4.05 $\times$ $10^{7}$ & 6.16 $\times$ $10^{11}$ & 7.60 $\times$ $10^{13}$\\
IB & (8,4) & 3.28 $\times$ $10^{5}$ & 4.99 $\times$ $10^{9}$ & 7.60 $\times$ $10^{13}$\\ 
IC & (6,2) & 4.05 $\times$ $10^{7}$ & -6.69 $\times$ $10^{12}$ & 6.99 $\times$ $10^{12}$\\
IC & (8,4) & 3.28 $\times$ $10^{5}$ & -4.83 $\times$ $10^{11}$ & 7.84 $\times$ $10^{11}$\\ \hline 
IIA & (6,2)&3.29$\times$ $10^{8}$& 9.73$\times$$10^{12}$&6.25$\times$$10^{16}$\\
IIA & (8,4)&2.63$\times$ $10^{6}$& 7.94$\times$$10^{10}$&6.21$\times$$10^{16}$\\
IIB & (6,2)&-9.97$\times$ $10^{8}$& 2.63$\times$$10^{12}$&5.59$\times$$10^{14}$\\
IIB & (8,4)&-8.10$\times$ $10^{6}$& 2.14$\times$$10^{10}$&5.57$\times$$10^{14}$\\
 \hline 
IIIA & (6,2)& 3.93$\times$ $10^{11}$&-4.09$\times$ $10^{11}$&2.87$\times$ $10^{14}$\\
IIIA & (8,4) &3.19$\times$ $10^{9}$&-3.22$\times$ $10^{9}$&2.85$\times$ $10^{14}$\\
IIIB & (6,2)& 3.85$\times$ $10^{11}$&-3.99$\times$ $10^{11}$&2.99$\times$ $10^{14}$\\
IIIB & (8,4) &3.13$\times$ $10^{9}$&-3.25$\times$ $10^{9}$&2.97$\times$ $10^{14}$\\
\hline
\end{tabular}
\hfil
\caption{\footnotesize  Heavy right-handed  Majorana neutrino  masses  $M_{j}$
 for  degenerate models (IA,IB,IC), inverted models (IIA,IIB)
and normal hierarchical models (IIIA, IIIB),  with
$\tan^2\theta_{12}$=0.5, using neutrino mass matrices given in
Appendix A. The entry $(m,n)$ in $m_{LR}$, indicates the type of Dirac neutrino
mass matrix taken as  charged lepton mass
matrix (6,2) or up quark mass matrix (8,4), as explained in the text.  }
\end{table}


\begin{table}[tbp]
\begin{tabular}{llllllll} 
\hline
Type & (m,n) & $\tilde{m_{1}}(GeV)$&${(h^{\dag}h)}_{11}$ & $k_{1}$ &$\epsilon_{1}$ &$Y_{B}$ \\
\hline
IA &(6,2)&1.19$\times 10^{-9}$&4.78$\times$ $10^{-6}$& 9.3$\times$$10^{-5}$&1.53$\times$$10^{-7}$&1.55$\times$$10^{-13}$\\
IA &(8,4)&1.19$\times 10^{-9}$&3.87$\times$ $10^{-8}$& 9.3$\times$$10^{-5}$&4.14$\times$$10^{-9}$&4.16$\times$$10^{-15}$\\
\hline
IB &(6,2)&3.97$\times 10^{-10}$&5.31$\times$ $10^{-7}$& 2.83$\times$$10^{-4}$&4.46$\times$$10^{-16}$&1.36$\times$$10^{-21}$\\
IB &(8,4)&3.97$\times 10^{-10}$&4.30$\times$ $10^{-9}$& 2.83$\times$$10^{-4}$&3.62$\times$$10^{-18}$&1.10$\times$$10^{-23}$\\
\hline
IC &(4,2)&3.97$\times 10^{-10}$&5.31$\times$ $10^{-7}$& 2.83$\times$$10^{-4}$&2.49$\times$$10^{-15}$&7.62$\times$$10^{-21}$\\
IC &(8,4)&3.97$\times 10^{-10}$&4.30$\times$ $10^{-9}$&
2.83$\times$$10^{-4}$&2.16$\times$$10^{-16}$&6.62$\times$$10^{-22}$\\
\hline 
IIA &(6,2)&4.95$\times 10^{-11}$&5.31$\times$ $10^{-7}$& 2.95$\times$$10^{-3}$&1.56$\times$$10^{-12}$&4.98$\times$$10^{-17}$\\
IIA &(8,4)&4.95$\times 10^{-11}$&4.30$\times$ $10^{-9}$& 2.95$\times$$10^{-3}$&1.26$\times$$10^{-14}$&4.04$\times$$10^{-19}$\\
\hline
IIB &(6,2)&1.08$\times 10^{-12}$& 5.01$\times$ $10^{-6}$& 8.83$\times$$10^{-4}$&2.69$\times$$10^{-7}$&2.57$\times$$10^{-12}$\\
IIB &(8,4)&1.52$\times 10^{-10}$&4.06$\times$ $10^{-8}$&
8.83$\times$$10^{-4}$&2.18$\times$$10^{-9}$&2.07$\times$$10^{-14}$\\
\hline 
IIIA &(6,2)&5.80$\times$ $10^{-10}$&7.51$\times$$10^{-3}$&1.82$\times$$10^{-4}$&4.59$\times$ $10^{-3}$&9.27$\times$ $10^{-9}$\\
IIIA &(8,4)&5.80$\times$ $10^{-10}$&6.13$\times$$10^{-5}$&1.82$\times$$10^{-4}$&3.62$\times$ $10^{-5}$&7.28$\times$ $10^{-11}$\\
\hline
IIIB &(6,2)&5.93$\times$ $10^{-10}$&7.51$\times$$10^{-3}$&1.83$\times$$10^{-4}$&4.91$\times$ $10^{-3}$&9.66$\times$ $10^{-9}$\\
IIIB &(8,4)&5.93$\times$ $10^{-10}$&6.13$\times$$10^{-5}$&1.83$\times$$10^{-4}$&3.88$\times$ $10^{-5}$&7.59$\times$ $10^{-11}$\\
\hline
\end{tabular}
\hfil
\caption{\footnotesize  Values of  CP asymmetry and  the baryon
 asymmetry for degenerate models (IA, IB, IC), inverted
 hierarchical models (IIA, IIB) and normal hierarchical models (IIIA,
 IIIB)  with for $\tan^{2}\theta_{12}$ =0.50, using mass matrices
 given in Appendix A. The entry $(m,n)$ indicates the type of Dirac
 mass matrix as explained in the text. }
\end{table}

\begin{table}[tbp]
 \begin{tabular}{c cll} \hline
Type & (m,n) &   $T^{min}_R<T_R\leq T^{max}_R$ (GeV)  &
$M^{min}_I<M_I\leq M^{max}_I$ (GeV)   \\ \hline
IA & (6,2) & $2.61\times 10^5 < T_R \leq 1.22\times 10^6$ & $2.44 \times 10^8 <M_I \leq 1.15 \times 10^9$    \\
IA & (8,4) & $7.80\times 10^4 < T_R \leq 9.86\times 10^3$ &
$1.97\times10^6<M_I\leq  2.49\times10^5$   \\ \hline
IB & (6,2) & $2.97\times 10^{13}<T_R \leq 4.05\times10^5$  &
$9.10\times10^7<M_I\leq 0.49$  \\
IB & (8,4) & $2.97\times10^{13}<T_R \leq  3.28\times10^3$
&$6.56\times10^5<M_I \leq  7.20\times10^{-5}$   \\ \hline
IC & (6,2) & $5.29\times10^{13}<T_R\leq 4.05\times10^5$ &
$8.05\times10^7<M_I \leq 2.74$ \\
IC & (8,4) & $4.97\times10^{11}<T_R \leq 3.28\times10^3$  &
$6.56\times10^5<M_I \leq 4.35\times10^{-3}$ \\ \hline 
IIA & (6,2)& $6.80\times10^{10}<T_R \leq 3.25\times10^6$  &
$6.50\times10^8<M_I \leq 3.09\times10^4$ \\
IIA & (8,4)& $6.80\times10^{10}<T_R \leq 2.64\times10^4$
&$5.26\times10^6<M_I \leq 2.03$ \\ \hline
IIB & (6,2)& $1.22\times10^6<T_R \leq 9.99\times10^6$  &
$1.99\times10^9<M_I \leq 1.64\times10^{10}$ \\
IIB & (8,4)& $1.22\times10^6<T_R \leq 8.1\times10^4$  &
$1.62\times10^7<M_I \leq 1.08\times10^6$  \\
 \hline 
IIIA & (6,2)& $2.80\times10^5<T_R \leq 3.93\times10^9$  &
$7.86\times10^{11}<M_I \leq 1.10\times10^{17}$  \\
IIIA & (8,4) & $2.88\times10^5<T_R \leq 3.19\times10^7$
&$6.38\times10^9<M_I \leq 7.06\times10^{12}$  \\ \hline
IIIB & (6,2)& $2.57\times10^4<T_R \leq 3.85\times10^9$  &
$7.70\times10^{11}<M_I \leq 1.15\times10^{17}$  \\
IIIB & (8,4) & $2.64\times10^4<T_R \leq 3.13\times10^7$  &
$6.26\times10^9<M_I \leq 7.42\times10^{12}$    \\
\hline
\end{tabular}
\hfil
\caption{\footnotesize Theoretical bounds on reheating temperature
  $T_R$ and inflaton mass $M_I$ in non-thermal leptogenesis, for all neutrino mass models described
in Tables 1-3.}
\end{table}
Our estimatated baryon asymmetry for normal hierarchical
model(IIIA, IIIB) lies between $9.27\times 10^{-9}$ with Dirac neutrino mass
matrix as charged lepton mass matrix $(6,2)$, and $7.28\times 10^{-11}$ for
the up-quark mass matrix $(8,4)$. This hints a possible choice of Dirac
neutrino mass matrix lying between these two e.g.,
$m_{LR}=diag.(\lambda^8, \lambda^2,1)v$. 
As emphasised earlier, our starting point is the neutrino mass matrix
which satisfies the observed  neutrino mass parameters and
mixings. The values of input parameters are fixed at this level before
applying to the calculation of baryon asymmetry. The whole calculation is performed
in a consistent way.

\section{Non-thermal leptogenesis}
We next apply  the neutrino mass models  discussed in section 2
(Tables 1-3) to  non-thermal
leptogenesis scenario [7] where the right-handed neutrinos are
produced through the direct non-thermal decay of the inflaton. We
follow the standard procedure oulined in ref. [20] where non-thermal leptogenesis and baryon
asymmetry in the universe had been  studied in different neutrino mass
models whereby  some mass models were excluded using bounds from below and
from above on the inflation mass and reheating temperature after
inflation. Though we adopt similar analysis, the texture of the neutrino mass
models considered here are different and hence the conclusions are also
expected to be different.

We start with the inflation decay rate given by 
\begin{equation}
\Gamma_{\phi}=\Gamma(\phi \rightarrow N_i N_i)\simeq
\frac{|\lambda_i|^2}{4\pi} M_I
\end{equation} 
where $\lambda_i$ are the Yukawa coupling constants for the
interaction of three heavy right-handed neutrinos $N_i$ with the
inflaton $\phi$ of mass $M_I$. The reheating temperature after
inflation   is given by the expression, 
\begin{equation}
T_R=\left(\frac{45}{2\pi^2 g_*}\right)^{1/4} (\Gamma_{\phi} M_P)^{1/2}
\end{equation}
where $M_{P}\simeq 2.4\times 10^{18}$ GeV is the reduced Planck mass[21]
and $g_*$ is the effective number of relativistioc degrees of freedom
at reheating temperature. For SM we have $g_*=106.75$ and for MSSM
$g_*=228.75$. If the inflaton dominantly couples to $N$, the branching
ratio of this decay process is taken as $BR\sim 1$, and the produced
baryon asymmetry of the universe can be calculated by the following
relation [22],
\begin{equation}
 Y_B=\frac{n_B}{s}=CY_L=C\frac{3}{2}\frac{T_R}{M_I}\epsilon
\end{equation}
where $Y_L$ is the lepton asymmetry generated by CP-violating
out-of-equilibrium decays of heavy neutrino $N_1$ and $T_R$ is the
reheating temperature. The fraction C has the value $C=-28/79$ for SM
and $C=-8/15$ in the MSSM.

 The above expression (12)  of the baryon asymmetry is
supplemented by two more boundary conditions [20]:
 (i) lower bound on inflaton mass $M_I>2 M_1$ coming from allowed
 kinematics of inflaton decay, and (ii) an upper bound for the
 reheating temperature  $T_R \leq 0.01 M_1$ coming from out-of-thermal
 equilibrium decay of  $N_1$. Using  the observed central value[2] of the
 baryon asymmetry  $Y_B= \frac{n_B}{s}=8.7\times 10^{-11}$ and theoretical prediction
 of CP asymmetry $\epsilon$ in Table 3, in  equation (12), one can
 establish the relation between  $T_R$ and $M_I$ for each neutrino
 mass model.

 The right-handed neutrino mass $M_1$ from Table 2 and the
 CP asymmetry $\epsilon$ from Table 3 for all neutrino mass models, are
 used to calculate the bounds:  $T^{min}_R < T_R \leq T^{max}_R$ and
 $M^{min}_I< M_I \leq M^{max}_I$  in Table 4 following eq.(12) along
 with other two boundary conditions cited above. Only those
 models which satisfy the constraint $T^{max}_R > T^{min}_R$ could
 survive in the non-thermal leptogenesis. These models are identified
 as  IA
 with  (6,2), IIB with (6,2), III (A,B) with (6,2) and III(A,B) with
 (8,4) where $(m,n)$ refers to the type of Dirac neutrino mass matrix.
  From Table 4 it is seen that inflationary models in which
 $M_I\sim 10^{13}$ GeV, like e.g., chaotic or natural inflation, are
 compatible only with normal hierarchical model III (A, B) with
 (6,2). In fact with $T_R =10^6$ GeV, we get $M_I= 2.8\times 10^{13}
 $ GeV., $\Gamma_{\phi}=2.85\times 10^{-6}$ GeV, and
 $|\lambda_1|=1.13\times 10^{-8}$ which are compatible with chaotic
 inflationary model. 

In supersymmetric models, the gravitino problem [8]  can be avoided provided
that the reheating temperature after inflation is bounded from above
in a certain way, namely $T_R\leq (10^6 - 10^7)$ GeV. In fact the
reheating
 temperature $T_R=10^6$ GeV is relevant  in
order to realise the weak scale gravitino mass $m_{3/2}\sim 100 $ GeV
without causing the gravitino problem. Even this reheating temperature
is relaxed for two order $T_R=10^7$ GeV, we would have $M_I\sim
10^{11}$ GeV in normal hierarchy type III(A,B) with (8,4).  We
conclude that the  only surviving model in this analysis  is the  normal hierarchical model
(III).  

\section{ Summary and discussion}
To summarise, we first parametrise the light left-handed Majorana
neutrino mass matrices describing the possible patterns of neutrino
masses viz, degenerate, inverted hierarchical and normal hierarchical,
which obey the $\mu-\tau$ symmetry having tribimaximal mixings. As a
first test these mass matrices predict the neutrino mass parameters
and mixings consistent with data, and all the input parameters are
fixed at this stage. In the next stage these mass matrices are
employed to estimate the baryon asymmetry in both thermal as well as
non-thermal leptogenesis scenario. We use the CP violating Majorana
phases derived from right-handed Majorana mass matrix and two possible
forms of Dirac neutrino mass matrices as either charged lepton mass
matrix or up-quark mass matrix in the calculation. The overall
analysis shows that normal hierarchical model appears to be the most
favourable choice in nature. The present analysis though
phenomenological may serve as an additional criteria to discard some
of the presently available neutrino mass models and neutrino mass
ordering patterns. There are some suggestions in the
literature[24] for inverted hierarchical model to enhance the
estimation of baryon asymmetry if $m_3$ is increased. The present
investigation has taken care of the maximum allowed  non-zero value of
$m_3\sim 0.033$ eV in case of inverted hierarchy type IIB model. Our result also
differs from a recent study in nonthermal leptogenesis with strongly
hierarchical right-handed neutrinos[25] where the mass of the lightest
right handed neutrino $M_1\leq 10^6$ GeV. There are some propositions[26]
for  probing the reheating
temperature at the Large Hadron Collider and this hopefully decides
the validity of thermal leptogenesis. 
\section*{Appendix A}
\subsection*{Classification:}
We first  list here for ready reference to the classification of
neutrino mass models,  the  zeroth-order left-handed Majorana neutrino mass 
matrices with texture zeros, $m_{LL}$, corresponding to three models
of neutrinos given in the text, viz., 
degenerate (Type [I]), inverted hierarchical (Type [II]) and normal hierarchical (Type [III]). 
\begin{center}
\begin{tabular}{ccc}\hline
Type  & $m_{LL}$ & $m_{LL}^{diag}$\\ \hline \\
\ [IA]    &${ \left(\begin{array}{ccc}
  0 & \frac{1}{\sqrt{2}} & \frac{1}{\sqrt{2}}\\ 
 \frac{1}{\sqrt{2}} & \frac{1}{2} & -\frac{1}{2}\\
 \frac{1}{\sqrt{2}} & -\frac{1}{2} & \frac{1}{2} 
\end{array}\right)}m_{0}$ & $Diag(1,-1,1)m_{0}$\\

\\
\ [IB]    &${ \left(\begin{array}{ccc}
  1 & 0 & 0\\ 
 0 & 1 & 0\\
 0 & 0 & 1 
\end{array}\right)}m_{0}$ & $Diag(1,1,1)m_{0}$\\
\\

\ [IC]    &${ \left(\begin{array}{ccc}
  1 & 0 & 0\\ 
 0 & 0 & 1\\
 0 & 1 & 0 
\end{array}\right)}m_{0}$ & $Diag(1,1,-1)m_{0}$\\ \hline
\\
\ [IIA]    &${ \left(\begin{array}{ccc}
  1 & 0 & 0\\ 
 0 & \frac{1}{2} & \frac{1}{2}\\
 0 & \frac{1}{2} & \frac{1}{2} 
\end{array}\right)}m_{0}$ & $Diag(1,1,0)m_{0}$\\
\\

\ [IIB]    &${ \left(\begin{array}{ccc}
  0 & 1 & 1\\ 
 1 & 0 & 0\\
 1 & 0 & 0 
\end{array}\right)}m_{0}$ & $Diag(1,-1,0)m_{0}$\\ \hline
\\

\ [III]    &${ \left(\begin{array}{ccc}
  0 & 0 & 0\\ 
 0 & \frac{1}{2} & -\frac{1}{2}\\
 0 & -\frac{1}{2} & \frac{1}{2} 
\end{array}\right)}m_{0}$ & $Diag(0,0,1)m_{0}$  \\ 

\\ \hline
\end{tabular}
\end{center}
\subsection*{Parametrisation with two parameters for tribimaximal mixings:}

 Left-handed Majorana neutrino mass matrices which obey
   $\mu-\tau $ symmetry[10,23] have the following form

 \[ m_{LL} = \left( \begin{array}{ccc}
 X  & Y  & Y \\
 Y &  Z & W \\
 Y &  W & Z \end{array} \right)m_{o} \]\\
This predicts an arbitrary solar mixing angle $\tan
2\theta_{12}=|\frac{2\sqrt{2}Y}{(X-Z-W)}|$, while the predictions on
atmospheric mixing angle is maximal $(\theta_{23}=\pi/4)$ and Chooz
angle zero. We parametrise the mass matrices (with only two parameters)
whereby the solar mixing is fixed at  tribimaximal mixings for all
possible patterns of neutrino mass models:

  1.\underline{Deg Type A [IA]}($m_i=m_1,-m_2,m_3$)\\
 \[ m_{LL} = \left( \begin{array}{ccc}
 \delta_{1}-2\delta_{2} & -\delta_{1}  & -\delta_{1} \\
-\delta_{1} &  \frac{1}{2}-\delta_{2} & -\frac{1}{2}-\delta_{2} \\
  -\delta_{1} &  -\frac{1}{2}-\delta_{2} & \frac{1}{2}-\delta_{2} \end{array} \right)m_{o} \]\\
with input values: $\delta_{1}$=0.66115, $\delta_{2}$=0.16535,$m_{o}=0.4 eV$.\\

  2.\underline{Deg Type B [IB]}($m_i=m_1,m_2,m_3$)\\
 \[ m_{LL} = \left( \begin{array}{ccc}
 1-\delta_{1}-2\delta_{2} & \delta_{1}  & \delta_{1} \\
\delta_{1} &  1-\delta_{2} & -\delta_{2} \\
  \delta_{1} & -\delta_{2} & 1-\delta_{2} \end{array} \right)m_{o} \]\\ 
with input values: $\delta_{1}$=8.314$\times$$10^{-5}$,$\delta_{2}$=0.00395,$m_{o}$=0.4eV.\\

 3.\underline{Deg Type C [IC]}($m_i=m_1,m_2,-m_3$)\\
 \[ m_{LL} = \left( \begin{array}{ccc}
 1-\delta_{1}-2\delta_{2} & \delta_{1}  & \delta_{1} \\
\delta_{1} &  -\delta_{2} & 1-\delta_{2} \\
  \delta_{1} & 1-\delta_{2} & -\delta_{2} \end{array} \right)m_{o} \]\\ 
with input values: $\delta_{1}$=8.314$\times$$10^{-5}$,$\delta_{2}$=0.00395,$m_{o}$=0.4eV.\\

4:\underline{Inverted Hierarchical mass matrix}
{\bf  with $m_3\neq 0$:}

\[ m_{LL}(IH)=
 \left(\begin{array}{ccc}
1 -2\epsilon  &  -\epsilon  &   -\epsilon \\
  -\epsilon   &  1/2  &   1/2-\eta  \\
 -\epsilon  &  1/2-\eta  &  1/2
 \end{array}\right)m_0.\] \\

 Inverted Hierarchy with even CP parity in the first two mass
 eigenvalues [IIA] $(m_1=m_1,m_2,m_3)$: $\eta /\epsilon$=1.0,$\eta$=0.0048,$m_{0}=0.05 eV$.\\
 Inverted Hierarchy with odd CP parity in the first two mass
 eigenvalues [IIB] $(m_i=m_1,-m_2,m_3)$: $\eta /\epsilon$=1.0,$\eta$=0.6607,$m_{0}=0.05 eV$.\\

5:.\underline{Normal Hierarchical mass matrix}
{\bf  Case (i)  with $m(1,1)\neq 0$} type- [IIIA]:

 \[m_{LL}(NH)=
 \left(\begin{array}{ccc}
 -\eta  &  -\epsilon  &   -\epsilon \\
  -\epsilon   &  1-\epsilon  &   -1  \\
 -\epsilon  &  -1  &  1-\epsilon
 \end{array}\right)m_0 \] \\
with input values: $\eta /\epsilon$=0.0,$\epsilon$=0.175,$m_{0}=0.029 eV$.\\

6:\underline{Normal Hierarchical mass matrix}
{\bf Case (ii)  with $m(1,1)=0$} type- [IIIB]:

\[ m_{LL}(NH)=
 \left(\begin{array}{ccc}
 0  &  -\epsilon  &   -\epsilon \\
  -\epsilon   &  1-\epsilon  &   -1+\eta  \\
 -\epsilon  &  -1+\eta  &  1-\epsilon
 \end{array}\right)m_0 \] \\
with input values: $\eta /\epsilon$=0.0,
$\epsilon$=0.164,$m_{0}=0.028eV$.

The textures of mass matrices for inverted hierarchy (IIA, IIB) as well
as normal hierarchy (IIIA, IIIB) have the potential to decrease the
solar mixing angle from the tribimaximal value, without sacrificing
$\mu-\tau$ symmetry. This is possible through the identification of
'flavour twister' $\eta/\epsilon \neq 0$ [23]. 



\begin{thebibliography}{50}
\bibitem{1}  M. Gell-Mann, P. Ramond and R. Slansky in Supergravity, Proceeding of the
Workshop, Stony Brook, New York, 1979, Edited by P. Van  Nieumenhuizen and 
 D. Freedman ( North-Holland, Amsterdam, 1979 ); T. Yanagida, KEK Lectures, 1979 (unpublished);
R. N. Mahapatra and G. Senjanovic, Phy. Rev. Lett. {\bf 44},912
(1980).
\bibitem{2} see for a recent review, Sacha Davidson, Enrico Nardi,
  Yosef Nir, {\bf arXiv:0802.2962} and further references therein.
\bibitem{3}  A. D. Sakharov, JETP Lett. {\bf 5} (1967) 24.

\bibitem{4}  V. A. Kuzmin, V. A. Rubakov, M. E. Shaposhnikov, Phy. Lett. {\bf 155B} (1985) 36.
\bibitem{5}  M. Fukugita and T. Yanagida, Phy.Lett. {\bf 174B}
  (1986) 45.
\bibitem{6} W. Buchmuller, P. Di Bari, M. Plumacher,
  Nucl. Phys. {\bf B643}, (2002)367, {\bf  hep-ph/0205349}; G. F. Giudice, A. Notari,
  M. Raidal, A. Riotto, A. Strumia, Nucl. Phys. {\bf B685}, (2004)89,
  {\bf hep-ph/0310123}.
\bibitem{7} G. Lazarides and Q. Shafi, Phys. Lett. {\bf B258},
  305(1991); K. Kumekawa, T. Moroi, T. Yanagida, Prog. Theor. Phys.{\bf 92},
  437(1994); G. F. Giudice, M. Peloso, A. Riotto and I. Tkachev, JHEP
  {\bf 9908}, 014 (1999) {\bf  arXiv: hep-ph/9905242}; T. Asaka, K. Hamaguchi,
  M. Kawasaki, T. Yanagida, Phys. Lett. {\bf B464}, 12(1999),
  Phys. Rev. {\bf D61}, 083512(2000); T. Asaka, H. B. Nielsen and
  Y. Takanishi, Nucl. Phys. {\bf B647}, 252(2002) {\bf arxiv: hep-ph/0207023};
  A. Mazumdar, Phys.Lett. {\bf  B580}, 7(2004) {\bf arXiv: hep-ph/0308020};
  T. Fukuyama, T. Kikuchi and T. Osaka, JCAP {\bf 0506}: 0506 (20050
  ,{\bf arXiv:
  hep-ph/0503201}.
\bibitem{8} J. R. Ellis, A.D. Linde, D.A. Nanopoulos,
  Phys. lett. {\bf B118}, (1982)59; M.Y. Khlopov, A. D. Linde,
  Phys. Lett. {\bf B138} (1984)265.  
\bibitem{9}  An incomplete list: P. F. Harrison, W. G. Scott, Phys. Lett. {\bf B547}, 219(2002);
  C. S. Lam, Phys. Rev. {\bf D71}, 093001(2005); {\bf hep-ph/0503157}; W. Grimus,
  {\bf hep-ph/0610158}; W. Grimus, L. Lavoura, J. Phys. G34: (2007)1757, {\bf  hep-ph/0611149};
  A. S. Joshipura, Eur. Phys. J. C53: (2008)77, {\bf hep-ph/0512252}; T. Kitabayashi, M. Yasue,
  Phys. Lett. {\bf B490}, 236(2000); E. Ma, Phys. Rev. {\bf D70}, 031901(2004);
  K. S. Babu, R. N. Mohapatra, Phys. Lett. {\bf B532}, 77(2002);
  T. Fukuyama, H. Nishiura, {\bf hep-ph/9702253}; K. Fuki, M. Yasue,
  Nucl. Phys. B783, (2007)31,
  {\bf hep-ph/0608042}; A. Ghosal, Mod. Phys. Lett. {\bf A19},
  2579(2004); {\bf hep-ph/0304090}; T. Ohlsson, G. Seidl,
  Nucl. Phys. {\bf B643}, 247(2002);  Riazuddin, Eur. Phys. J. C51, (2007)699,
  {\bf arXiv:0707.0912}; Takeshi Fukuyama, arXiv:0804.2107;  Y. H. Ahn, Sin Kyu Kang, C. S. Kim, Jake Lee,
  Phys.Rev.D73: (2006)093005, {\bf hep-ph/0602160};  Yoshio Koide, Phys. Rev. {\bf D69}, 093001(2004); Yoshio Koide,
  H. Nishiura, K. Matsuda, T. Kikuchi, T. Fukuyama, Phys. Rev. {\bf D66},
  093006(2002); Koichi Matsuda, H. Nishiura, Phys. Rev. {\bf D73},
  013008(2006);  Y. Koide, E. Takasugi, Phys. Rev. {\bf D77}, (2008)016006,
  {\bf arXiv:0706.4373};  R. N. Mohapatra, S. Nasri, Hai-Bo Yu, Phys. Lett. {\bf B636},
  114(2006).

\bibitem{10} P. F. Harrison, D. H. Perkin, W. G. Scott,
  Phys. Lett. {\bf B530}, 167(2002), {\bf hep-ph/0202074};
  P. F. harrison, W. G. Scott, Phys. Lett. {\bf B557}, (2003)76.
\bibitem{11}  M. A. Luty, Phy.Rev. {\bf D45} (1992) 455.
\bibitem{12}  E. W. Kolb, M. S. Turner, \textsl{The Early
    Universe}, Addision - Wesely, New York (1990).
\bibitem{13} M. Flanz, E. A. Paschos, U. Sarkar, Phys. Lett. {\bf B345}, 248
  (1995) [ Erratum-ibid. {\bf B382}, 447(1996)]; L. Covi, E. Roulet and
  F. Vissani, Phys. Lett. {\bf B384}, 169(1996); M. flanz, E. A. Paschos,
  U. Sarkar and J. Weiss, Phys. Lett. {\bf B389}, 693(1996); A. Pilaftis,
  Phys. Rev. {\bf D56}, 5431 (1997); W. Buchmuller and M. Plumacher,
  Phys. Lett. {\bf B431}, 354 (1998). 
\bibitem{14} A. Pilaftsis, Phys.Rev. {\bf D56}, 5431 (1997);
  A. Pilaftis, T. E. J. Underwood, Nucl.Phys. {\bf B692}, 303(2004).
\bibitem{15} S. Y. Khlebnikov, M.E.Shaposhnikov, Nucl. Phys. {\bf B308},
  885(1988).  W. Buchmuller, R.D. Peccei and T. Yanagida, {\bf arxiv:
    hep-ph/0502169}, W. Buchmuller, {\bf arXiv: 0710.5857}.
\bibitem{16} P. Di Bari, {\bf  hep-ph/0406115}, {\bf hep-ph/0211175};
  W. Buchmuller, P. Di. Bari, M.Plumacher, Nucl. Phys. {\bf B665},
  445(2003).
\bibitem{17} E.K.Akhmedov, M. Frigerio and A. Y. Smirnov, JHEP
  {\bf 0309}, 021(2003),{\bf  hep-ph/0305322}.
\bibitem{18}  A.K.Sarma, H.Z.Devi and N.Nimai Singh, Nucl.Phys.{\bf B765}(2007)142-153.
\bibitem{19}  K. S. Babu, A. Bachri and  H. Aissaoui,Nucl.Phys.{\bf
    B738}(2006) 76-92, {\bf hep-ph/0509091}. 
\bibitem{20} G. Panotopoulos, Phys. Lett. {\bf B643} (2006)279.
\bibitem{21} Frank Daniel Steffen, {\bf arXiv: 0806.3266}.
\bibitem{22} W. Buchmuller, R. D. Peccei, T. Yanagida, Annual
  Rev. Nucl. Part. Sci. {\bf 55} (2005)311, {\bf hep-ph/0502169}.
\bibitem{23} N.Nimai Singh, H. Zeen Devi, Mahadev Patgiri, {\bf arXiv:
    0707.2713}; N. Nimai Singh, M. Rajkhowa, A. Borah, J. Phys. G:
  Nucl. Part. Phys.{\bf  34}, (2007)345; Pramana J. Phys. {\bf 69}, (2007)533. 
\bibitem{24} E.Molinaro, S.T. Petcov, T. Shindou, T. Takanishi, {\bf arXiv:
  0709.0413}.
\bibitem{25} V. Nefer Senoguz, Phys. Rev.{\bf  D76}, (2007) 013005; {\bf arXiv:
  0704.3048}.
\bibitem{26} Frank Daniel Steffen, {\bf arXiv: 0806.3266}. 

 \end{thebibliography}
\end{document}